\documentclass[%
  aps,prd,
  12pt, tightenlines,
  nofootinbib,preprintnumbers,showpacs, 
  amsmath,amssymb]{revtex4}
\usepackage{mathrsfs}
\allowdisplaybreaks

\newcommand{\dd}{\mathrm{d}}
\newcommand{\ii}{\mathrm{i}}
\newcommand{\dmu}{\partial_\mu}
\makeatletter
\newcommand{\fmslash}[2][0mu]{%
  \mathchoice
    {\fmsl@sh\displaystyle{#1}{#2}}%
    {\fmsl@sh\textstyle{#1}{#2}}%
    {\fmsl@sh\scriptstyle{#1}{#2}}%
    {\fmsl@sh\scriptscriptstyle{#1}{#2}}}
\newcommand{\fmsl@sh}[3]{%
  \m@th\ooalign{$\hfil#1\mkern#2/\hfil$\crcr$#1#3$}}
\makeatother
\begin{document}
\preprint{MZ-TH/04-01}
\preprint{hep-ph/0402118}
\title{Higgsless fermion masses and unitarity}
\author{Christian Schwinn}
\email{schwinn@thep.physik.uni-mainz.de}
\affiliation{Institut f\"ur Physik, Johannes-Gutenberg-Universit\"at,
 Staudingerweg 7,  D-55099 Mainz, Germany}
\begin{abstract}
We discuss the consistency of fermion mass generation by boundary conditions
and brane localized terms 
in higher dimensional Higgsless models of gauge symmetry breaking.
The sum rules imposed by tree-level unitarity and Ward Identities
are applied to check the
consistency of mass generation  by 
orbifold projections and more general boundary conditions consistent
with the variational principle.
We find that the sum rules are satisfied for
boundary conditions corresponding to brane localized mass 
and kinetic terms consistent with the reduced
gauge symmetry on the brane. 
\end{abstract}
\maketitle
\section{Introduction}
Recently a new class  of models of
electroweak symmetry breaking (EWSB) without Higgs bosons 
has been proposed~\cite{Csaki:2003dt} in the setting of an additional
dimension at the TeV
scale~\cite{Antoniadis:1990ew}.
Since Higgsless EWSB 
is not possible using the abelian 
orbifold constructions
 usually employed in higher dimensional models of grand unified 
theories~(GUTs)~\cite{Kawamura:1999nj,Hall:2001tn,Hebecker:2001jb},
 a more general approach to 
symmetry breaking by Dirichlet boundary conditions~(BCs)
has been utilized in these models.  
This construction has been found to be 
consistent with the variational principle,
tree level unitarity~\cite{Csaki:2003dt} 
and Becchi-Rouet-Stora-Tyutin~(BRST) symmetry~\cite{OS:SR}.

There are by now several variants of this setup in warped 
space~\cite{Csaki:2003zu,Davoudiasl:2003me,Cacciapaglia:2004jz}
and in flat space~\cite{Barbieri:2003pr}.
While there is some tension in satisfying 
precision data and constraints
from partial wave unitarity 
at the same time~\cite{Barbieri:2003pr,Davoudiasl:2003me}, 
it has been suggested that these problems can be overcome by  appropriate
brane kinetic terms~(BKTs) for the gauge bosons~\cite{Cacciapaglia:2004jz}. 

In the Higgsless higher dimensional models, the unitarity sum rules~(SRs) 
that guarantee the boundedness of the amplitude
at large energies are satisfied
because of interlacing cancellations among the Kaluza-Klein~(KK)-states 
of the gauge 
bosons~\cite{Csaki:2003dt,SekharChivukula:2001hz} instead of
the exchange of a scalar boson like in 4 dimensional dimensional 
theories~\cite{LlewellynSmith:1973,Lee:1977}.
This observation has also inspired new four dimensional 
models of Higgsless EWSB~\cite{Foadi:2003xa}
with improved unitarity properties.
While the most dramatic violations of unitarity can occur in the scattering
of massive gauge bosons, in the Standard Model~(SM)
the Higgs mechanism is also invoked to cancel divergences 
in the amplitudes for gauge boson production by fermions. 
Therefore, without a Higgs boson, also an inherently higher dimensional 
mechanism has to be employed to generate fermion masses without
spoiling unitarity.
In~\cite{Csaki:2003sh} bulk fermions with BCs corresponding 
to brane localized mass terms  
have been proposed for that purpose.
Such mass terms have been discussed for 
Majorana fermions in the context
of M-theory or Supergravity breaking~\cite{Antoniadis:1997ic,Bagger:2001qi,
Meissner:2002dg,Delgado:2002xf}
or Neutrino masses~\cite{Huber:2003sf}, but
brane induced  Dirac masses and the corresponding
KK-decomposition in the context of gauge symmetry
breaking have been discussed in less detail.

The consistency of the BCs corresponding to boundary mass terms with
 the variational principle has been discussed 
in~\cite{Csaki:2003sh} but in gauge theories also 
the interplay of fermion and gauge boson BCs has to be taken into account in
order not to violate unitarity or the Ward identities~(WIs) resulting from BRST
symmetry.
Since gauge symmetry breaking by orbifold and Dirichlet BCs leads to a reduced 
gauge symmetry on the boundary~\cite{Hall:2001tn}, 
the inclusion of brane localized terms invariant
under the reduced gauge symmetry can be regarded
as an explicit but `soft' symmetry breaking. 
For brane localized matter fields, the consistency of this explicit breaking
 with unitarity and WIs has been checked in 
\cite{Hall:2001tn,OS:SR} and in this work
we extend this analysis to  brane localized mass and
 kinetic terms of bulk fermions.

In section~\ref{sec:higgsless} we review briefly the main ingredients of 
Higgsless EWSB and fermion masses. 
In section~\ref{sec:wi-results} we 
verify the SRs from tree unitarity and  WIs for fermion masses resulting
from orbifold projections and discuss the possibility for
Higgsless fermion masses in this framework. 
In section~\ref{sec:brane-masses}, the consistency of 
fermion masses from brane localized mass and kinetic terms and
from mixing with brane fermions is discussed.

\section{Higgsless EWSB and fermion masses}\label{sec:higgsless} 
We now outline the mechanism of  
Higgsless EWSB by Dirichlet BCs and the generation of fermion masses by 
boundary terms in the flat space toy model of~\cite{Csaki:2003dt,Csaki:2003sh}.
This is not yet a viable model and to obtain the correct masses 
for the weak gauge bosons,
one must either introduce a warped
compactification~\cite{Csaki:2003zu} or include
 brane localized kinetic terms~\cite{Barbieri:2003pr}. 
However for
the issue of fermion mass generation, the essential features
can be discussed already in the simpler framework provided by the
flat space toy model.
\subsection{Higgsless gauge symmetry breaking}
The mechanism of Higgsless EWSB proposed in~\cite{Csaki:2003dt} 
prescribes Dirichlet BCs to the gauge bosons associated to the
broken symmetry generators (identified by a hat) and
Neumann BCs to the unbroken gauge bosons:
\begin{equation}
\label{eq:dirichlet}
  A^{\hat a}_\mu(y_f)=0\quad,\quad \partial_y A_\mu^{a}(y_f)=0
\end{equation}
where $y_f$ denotes one of the endpoints of the interval $[0,\pi R]$ of
the fifth dimension.
This symmetry breaking by BCs
allows to avoid the group theoretical constraints from abelian orbifold 
symmetry breaking~\cite{Hebecker:2001jb}. 
Physically, the Dirichlet BCs
can arise from the coupling to a boundary Higgs boson with a vacuum
 expectation value that is
pushed to infinity~\cite{Csaki:2003dt,Hebecker:2001jb}.
The important common property of Dirichlet BCs
and orbifold breaking ensuring unitarity 
of gauge boson scattering~\cite{Csaki:2003dt} and BRST 
symmetry~\cite{OS:SR}
is the vanishing of the wavefunctions of the broken gauge
bosons at the branes. Via the gauge transformation law, 
this implies also the vanishing of the gauge parameters corresponding
to the broken generators (and the vanishing of the wavefunctions
 of the ghosts associated to the broken gauge bosons in the BRST formalism~\cite{OS:SR}). Therefore the concept
of a reduced gauge or BRST symmetry on the branes can be introduced similar 
to the orbifold case~\cite{Hall:2001tn}. 

The group structure of Higgsless EWSB 
employed in~\cite{Csaki:2003dt} is
a left-right symmetric bulk symmetry group
$SU(2)_L\times SU(2)_R\times U(1)$.
On the brane at $y=\pi R\equiv\ell$ the left-right symmetry
is broken to the diagonal subgroup $SU(2)_{L+R}$ 
by assigning  Dirichlet BCs 
to the broken components  
$A^{-,a}_\mu\equiv\frac{1}{\sqrt 2}(A^{L,a}_\mu-A^{R,a}_\mu)$ while
the unbroken components 
$A^{+,a}_\mu\equiv\frac{1}{\sqrt 2}(A^{L,a}_\mu+A^{R,a}_\mu)$ 
and  $B_\mu$ satisfy Neumann BCs.
Analogously, on the second brane at $y=0$
the symmetry is broken according to $SU(2)_R\times U(1)\to U(1)_Y$
by prescribing Dirichlet BCs to the fields
    $A^{R,1/2}_\mu$ and $(g'B_\mu-gA_\mu^{R,3})$ 
while the unbroken fields are  
    $A_\mu^{L,a}$ and the linear combination $(gB_\mu+g'A_\mu^{R, 3})$.
The only unbroken component that satisfies Neumann BCs at both branes and
 therefore possesses a massless zero mode is the
combination $\gamma_\mu:=gB_\mu+g'(A_\mu^{R,3}+A_\mu^{L,3})$.

Using a theory space approach, a simpler setup involving only a bulk $SU(2)$
has been proposed~\cite{Foadi:2003xa} but a prescription 
how to incorporate fermions in a true five dimensional version of the model
has not been given. 
\subsection{Higgsless fermion masses}
To reproduce the SM fermion spectrum, in~\cite{Csaki:2003sh} 
 a bulk fermion is  introduced for every chiral fermion.
As an example, the left handed doublet $(u_L,d_L)$ is the zero mode of a
bulk $SU(2)_L$ doublet while the righthanded quarks $u_R, d_R$ are
 contained in a bulk $SU(2)_R$ 
doublet\footnote{Our notation for fermions in five dimensions
is introduced in section~\ref{sec:orbifolds} 
and appendix~\ref{app:kk-ferm}} 
\begin{equation}
\begin{aligned}
\Psi_{Q_L}&=\left(\Psi_{u_L},\Psi_{d_L}\right)=
\left(\begin{pmatrix}u_L\\\eta_{u_L}\end{pmatrix},
\begin{pmatrix}d_L\\\eta_{d_L}\end{pmatrix}\right)\\
\Psi_{Q_R}&=\left(\Psi_{u_R},\Psi_{d_R}\right)=
\left(\begin{pmatrix}\chi_{u_R}\\u_R\end{pmatrix},
\begin{pmatrix}\chi_{d_R}\\d_R\end{pmatrix}\right)
\end{aligned}
\end{equation}
The righthanded massless modes of $\Psi_{Q_L}$ and the lefthanded massless 
modes of $\Psi_{Q_R}$ can be projected out by BCs similar to those of fermions
on an orbifold~\cite{Georgi:2000,Papavassiliou:2001be}~(see
section~\ref{sec:orbifolds}). 

In~\cite{Csaki:2003sh}, brane localized mass terms are used to 
generate Dirac masses for the zero modes of bulk fermions.
On the brane at $y=\ell$,  only 
 the diagonal subgroup of $SU(2)_L\times SU(2)_R$ is
unbroken and Dirac masses terms consistent with this symmetry can be added:
\begin{equation}\label{eq:dirac-brane}
\mathscr{L}_{\ell}=-\delta(y-\ell)M_D\ell( \bar \Psi_L\Psi_R+\bar \Psi_R\Psi_L)
\end{equation}
On the brane at $y=0$, the broken $SU(2)_R$ allows to lift the degeneracy 
of the masses in the isospin multiplets.
A mass splitting in the quark sector can be generated by 
introducing brane localized  vectorlike 
fermions $\psi_u=(\chi_u, \eta_u)$ that have the same quantum numbers 
as the up-type quarks. 
One can then add a brane term mixing the right handed bulk up-type quarks
 with the brane fermions
\begin{equation}\label{eq:brane-mix}
\mathscr{L}_{0}=-\delta(y)M\ell^{1/2} 
(\bar u_R\chi_u+\bar \chi_u u_R)
\end{equation}
to generate a mass splitting among the up-and down type quarks.
The mass splitting in the lepton doublet can be generated analogously.
As noted in~\cite{Csaki:2003sh}, a similar mass spectrum 
can be obtained from BKTs~\cite{Dvali:2001gm} for the 
right handed bulk fermions
without the need to introduce brane localized fermions.

As in the model of Nomura in~\cite{Csaki:2003zu}
there are other  less minimal possibilities to obtain the SM fermion spectrum,
employing the same ingredients of brane mass terms and 
mixing with brane fermions.
In the remainder of this work, we consider generic features of BKTs and  
mass terms of the form~\eqref{eq:dirac-brane}
and~\eqref{eq:brane-mix} so our conclusions apply to any construction
using these ingredients.
This should also be useful
in a more general context than EWSB like Higgsless breaking of 
GUT symmetries that is not possible from orbifolding, e.g. the breaking of
 $SO(10)$ to the SM in five dimensions.
\section{Consistency of fermion masses from orbifolding}
\label{sec:wi-results}
Since the symmetry breaking $SU(2)_L\times SU(2)_R\to SU(2)_{L+R}$
 used in Higgsless EWSB is isomorphic  
to the allowed orbifold symmetry breaking
pattern~\cite{Hebecker:2001jb} $SO(4)\to SO(3)$, it 
can also be described in terms of orbifold breaking, 
as already noted in~\cite{Csaki:2003dt}.
In section~\ref{sec:orbifolds} 
we discuss the generation of fermion masses 
from this perspective, providing an alternative to the brane localized
mass terms. 
We then apply the methods of~\cite{OS:SR} to bulk fermions
and verify the consistency of gauge symmetry breaking by orbifold BCs. 
The extension to the more general BCs resulting from brane localized terms
is discussed in section~\ref{sec:brane-masses}.

\subsection{Orbifold symmetry breaking and Higgsless fermion masses}
\label{sec:orbifolds} 
Before we turn to the orbifold description of the symmetry breaking
$SU(2)_L\times SU(2)_R\to SU(2)_{L+R}$ we need to recall some 
results on fermions on a 5 dimensional
 orbifold~\cite{Georgi:2000,Papavassiliou:2001be}
in the connection of 
gauge 
symmetry breaking 
(see e.g.~\cite{Hebecker:2001jb,vonGersdorff:2002as}).
 The orbifold $S/(Z_2\times Z_2')$ is obtained from the circle by identifying
points under the action of the two reflections
   $ Z_2: y\to -y$ and 
    $Z_2':(y-\pi R)\to-(y-\pi R)$.
Fields defined on an orbifold
need only be invariant under the reflections
up to transformations $\mathcal{Z}$ that are a representation of $Z_2$ in
field space and leave the lagrangian invariant.
In the following, gauge bosons in five dimensions are decomposed into
a four dimensional vector and a scalar according to\footnote{
See appendix~\ref{app:kk} for our
 conventions for the KK decompositions and the resulting effective
lagrangian of the KK-modes.} 
$A^a_M=(A^a_\mu,-\phi^a)$.
For the case of a five dimensional gauge theory coupled to fermions, 
the  combined orbifold 
transformations take the form~\cite{vonGersdorff:2002as}
\begin{equation}\label{eq:orbi-trans}
\begin{aligned}
A^a_\mu(x,-(y-y_f)) )&=\mathcal{Z}^{y_f}_{ab}A^b_\mu(x,y-y_f)\\
\phi^a(x,-(y-y_f))&=-\mathcal{Z}_{ab}^{y_f}\phi^b(x,y-y_f)\\
\Psi_i(x,-(y-y_f))&=\gamma^5 \lambda^{y_f}_{ij} \Psi_j(x,y-y_f)
\end{aligned}
\end{equation}
with $y_f=0,\ell=\pi R$.
The  representation matrices $\mathcal{Z}^{y_f}_{ab}$ must satisfy
$\mathcal{Z}^2=1$ so their eigenvalues  are $\eta_a^{y_f}=\pm 1$.
For $\eta_a^{y_f}\neq 1$, the wavefunctions of the vector components of the
gauge fields must vanish at the 
fixed point $y_f$ so there the gauge symmetry is broken to a subgroup $H$.
The $\lambda$ are hermitian matrices 
 acting in the representation of the fermions, satisfying in addition
 $\lambda={\lambda}^{-1}$. 
In order  to leave the interaction with the 
gauge bosons invariant
under the transformation~\eqref{eq:orbi-trans}, 
in the basis where the $\mathcal{Z}$ are 
diagonal, the condition 
\begin{equation}\label{eq:fermion-cons}
\lambda \tau^a \lambda=\eta_a \tau^a
\end{equation}
must be satisfied. Here the $\tau$ are the generators of the gauge group
in the representation of the fermions.

The KK-decomposition of five dimensional fermions
is introduced as (c.f. appendix~\ref{app:kk-ferm} for details)
\begin{equation}\label{eqref:fermi-kk}
\Psi_i(x,y)=\sum_{n}\begin{pmatrix}\chi_{i,n}(x)\zeta^-_{i,n}(y)\\
                                   \eta_{i,n}(x)\zeta^+_{i,n}(y)\end{pmatrix}
\end{equation}
where the 4-dimensional spinors of the KK modes 
satisfy the Dirac equation~\eqref{eq:4d-dirac}.

In the basis where the ${\lambda^{y_f}}$ 
are diagonal with eigenvalues $\lambda_i^{y_f}=\pm 1$, 
the KK modes transform under the orbifold transformation according 
to~\cite{Georgi:2000,Papavassiliou:2001be}
\begin{equation}\label{eq:fermion-orbi}
\zeta^{\pm}_{i,n}(-(y-y_f))=\pm\lambda_i^{y_f}\zeta^{\pm}_{i,n}(y-y_f)
\end{equation}
Therefore for $\lambda_i=1$ only the 
right-handed fermions possess zero-modes, for $\lambda_i=-1$ only the left-handed. 

We will now  introduce an orbifold approach to 
the breaking $SU(2)_L\times SU(2)_R\to SU(2)_{L+R}$
and discuss how Dirac masses
for the lightest fermions can be obtained.
We collect the gauge fields in a vector $A_M=(A_L,A_R)$ and assign
the orbifold parities 
\begin{equation}\label{eq:outer-auto}
\mathcal{Z}^0=\mathbf{1}\quad,\quad \mathcal{Z}^{\ell}=\sigma^1  
\end{equation}
where  the sigma matrices act on the $L/R$ indices,
not on the indices of the gauge group.
The transformation~\eqref{eq:outer-auto} is diagonalized by going to the basis
\begin{equation}
    A_M^{\pm,a} =\frac{1}{\sqrt 2}(A_M^{L,a}\pm A_M^{R,a})
\end{equation}
so that $\eta_\pm^\ell =\pm 1$.

As a simple model for the setup discussed 
in section~\ref{sec:higgsless}, we will consider two bulk fermion doublets:
 $\Psi_L$ charged under $SU(2)_L$ and 
$\Psi_R$ charged under $SU(2)_R$.
Assembling $\Psi_L$ and 
$\Psi_R$ in a vector $\Psi=(\Psi_L,\Psi_R)$, the
interaction  lagrangian of the fermions can be written as 
\begin{equation}\label{eq:lr-lag}
\mathscr{L}_{f,\text{int}}=
\frac{1}{\sqrt 2}\bar \Psi_i(x,y) \tau^a_{ij}\Gamma^M  \left[A^{+,a}_M(x,y)+\sigma_3
  A^{-,a}_M(x,y)\right]  \Psi_j(x,y) 
\end{equation}
In contrast to the example of five dimensional QED~\cite{Papavassiliou:2001be},
the left-right symmetric gauge symmetries in the bulk
 forbid the addition of explicit 
bulk mass terms connecting $\Psi_R$ and $\Psi_L$.
Instead, we can give mass to the lightest KK modes using
 an appropriate orbifold transformation.
The generators of the  transformation of the fermions 
must satisfy~\eqref{eq:fermion-cons},
 i.e. $\lambda^\ell$ has to  anticommute with $\sigma^3$. 
To mix $\Psi_L$ and $\Psi_R$ we choose
\begin{equation}
  \lambda^{0}=-\sigma_3 \quad,\quad \lambda^{\ell}=\sigma_1
\end{equation}
where again the sigma matrices act on the $L/R$ indices.
More explicitly, the transformations of the fermions are given by
\begin{equation}\label{eq:iso-orbi}
  \begin{aligned}
    \Psi_{L/R}(x,-y)&=\mp \gamma^5 \Psi_{L/R}(x,y) \\
    \Psi_L(x,-(y-\ell)&=\gamma^5\Psi_R(x,y-\ell)
  \end{aligned}
\end{equation}
so at $y=0$ we project out the righthanded zero modes of $\Psi_L$ 
and the lefthanded zero modes of   $\Psi_R$ while the transformation
at $y=\ell $ mixes $\Psi_L$ and $\Psi_R$ in order to generate Dirac masses
for the surviving zero modes.
The orbifold symmetries~\eqref{eq:iso-orbi} are consistent with a bulk mass
term $m(y)(\bar\Psi\Psi)$ where $m(y)$ is an odd function under the orbifold
transformations.

As a simple example, we consider a mass term  given by a
step function $m(y)=m\epsilon(y)$. The KK-wavefunctions for more usual 
BCs have been given e.g. in~\cite{Csaki:2003sh}.
Imposing the orbifold condition~\eqref{eq:iso-orbi}  at $y=\ell$
we obtain the
KK wavefunctions in the interval $[0;\ell]$ up to normalization as
\begin{equation}\label{eq:kk-wfs-twist}
  \begin{aligned}
    \zeta^-_{L,n}(y)&=\left(\cos(k_n y)
      +\tfrac{m}{k_n}\sin(k_n y)\right)\\
    \zeta^+_{L,n}(y)&=\tfrac{m_n}{k_n}\sin(k_n y)\\
    \zeta^-_{R,n}(y)&=-\beta_n\tfrac{m_n}{k_n}\sin(k_n y)\\
     \zeta^+_{R,n}(y)&=\beta_{n}\left(\cos(k_n y)
      -\tfrac{m}{k_n}\sin(k_n y)\right)
  \end{aligned}
\end{equation}
with $\beta_n m_n=m\pm\sqrt{m^2+m_n^2}$ and  $k_n^2=m_n^2-m^2$.
The quantization condition for the masses is given by:
\begin{equation}\label{eq:orbi-masses}
  \begin{aligned}
  &  &\zeta^-_{L,n}(\pi R)&=-\zeta^-_{R,n}(\pi R)\\
  &\Leftrightarrow&  \tan(k_n\pi R)&=\pm\frac{k_n}{\sqrt{m^2+m_n^2}}
  \end{aligned}
\end{equation}
Therefore, our construction indeed gives Dirac masses connecting 
the left-handed component of $\Psi_L$ and the right handed component
of $\Psi_R$ that are determined by the bulk mass $m$.
It is beyond the scope of this paper to discuss the viability of this setup
in the context of a realistic model.
At any rate, the split of the masses within the isospin multiplets 
has to occur in the context of the $SU(2)_R\times U(1)\to U(1)_Y$ breaking
at $y=0$ that cannot be achieved by orbifolding alone.
\subsection{Consistency with unitarity and WIs}
\label{sec:orbi-sr}
Having described how to obtain fermion masses from orbifold BCs, 
we now turn to the verification of the consistency of orbifold
 symmetry breaking, 
using tree level unitarity and WIs as criteria.
In KK-gauge theories, using an appropriate gauge fixing, 
 WIs similar to those in a four dimensional spontaneously broken gauge 
theory~(SBGT) 
are valid with the scalar component of the five dimensional gauge boson taking
the role of the Goldstone bosons~(GBs)~\cite{SekharChivukula:2001hz,OS:SR}. 
In \cite{OS:SR} it has been shown that the SRs ensuring 
unitarity cancellations can also be derived by imposing those WIs
on a minimal set of scattering amplitudes.
Thus  the unitarity conditions for the fermion 
couplings \cite{LlewellynSmith:1973}~can be obtained  in a simpler way
from the WIs for the  $\bar f f\to WW$  amplitude.

In gauge boson production from 
fermions, the cancellation of the terms growing with the square of the energy  
is ensured by the relation~(see~\eqref{eq:kk-couplings} 
and~\eqref{eq:ferm-couplings} for the definition of the coupling constants)
\begin{subequations}\label{eq:fermi-srs}
\begin{equation}\label{eq:fermi-lie}
\mathcal{T}_{R/LIJ}^\alpha\mathcal{T}_{R/LJK}^\beta-\mathcal{T}_{R/LIJ}^\beta\mathcal{T}_{R/LJK}^\alpha
=g^{\alpha\beta\gamma}\mathcal{T}_{R/LIK}^\gamma
\end{equation}
 that has the form of a Lie algebra for the $\mathcal{T}_{R/L}$.
Here we have combined the KK and the group indices into multi-indices
 $(n,i)\equiv I$ and $(a,i)\equiv\alpha$ and have used a summation convention.
The cancellation of the subleading divergences $\propto E$ implies the relation
\begin{equation}\label{eq:fermi-gold-KK}
\ii g_{L IJ}^\beta\mathcal{T}_{LJK}^\alpha  -\ii \mathcal{T}_{RIJ}^\alpha
g_{L JK}^\beta= g_{L IK}^\gamma T^\alpha_{\beta\gamma}
\end{equation}
\end{subequations}
This relation can be interpreted as an invariance condition of the
Yukawa coupling $ g_{L IJ}^\beta$ under transformations generated by the
$T$ and $\mathcal{T}$.
In general, an additional term involving the Higgs coupling appears in
 this condition~\cite{LlewellynSmith:1973,OS:SR}.
In~\eqref{eq:fermi-gold-KK}
 the coupling of the fermions to the GBs $g_{L IK}^\gamma$ has to 
satisfy the relation
\begin{subequations}\label{eq:3-wis}
 \begin{equation}\label{eq:f-phi-coupling}
m_\alpha g_{L IJ}^\alpha 
=-\ii( m_I\mathcal{T}_{LIJ}^\alpha -m_J\mathcal{T}_{RIJ}^\alpha)
\end{equation}
and the coupling of the GBs to the gauge bosons has to satisfy
\begin{equation}\label{eq:2phi-w-coupling}
 T^{\alpha}_{\beta\gamma}=
\frac{1}{2m_\beta m_\gamma}g^{\alpha\beta\gamma}
(m_\alpha^2-m_\beta^2-m_\gamma^2)
\end{equation}
\end{subequations}
These conditions arise from the WIs for three point vertices~\cite{OS:SR}.

The fulfillment of the SRs~\eqref{eq:fermi-srs} and~\eqref{eq:3-wis}
 is a necessary but no sufficient condition
for perturbative unitarity. 
In four dimensional 
SBGTs, an upper bound on the Higgs mass can be obtained
demanding that unitarity cancellations 
set in before partial wave unitarity gets violated ~\cite{Lee:1977}.
Although a rigorous derivation has not yet been given, similar considerations
in Higgsless higher dimensional models
lead to an upper bound on the masses of the lightest KK excitations of the
gauge bosons as discussed by 
Davoudiasl~\emph{et.al} in~\cite{Davoudiasl:2003me}.
Furthermore, partial wave unitarity in compactified 
higher dimensional  gauge theories is violated by the infinite number
of KK modes so these theories have to be considered as effective
theories valid below a scale determined by the higher dimensional
dimensionfull gauge coupling constant~\cite{SekharChivukula:2001hz}.
In the following, we will only be concerned with the SRs~\eqref{eq:fermi-srs}
and~\eqref{eq:3-wis} ensuring the cancellation
of the terms diverging with the energy.

To check~\eqref{eq:f-phi-coupling}, using the relation among the gauge boson 
wavefunctions~\eqref{eq:wf-relation} and the equations of 
motion~\eqref{eq:ferm-kk-eom} we obtain, integrating by parts:
\begin{align}\label{eq:f-phi-orbi}
m_\alpha g_{L IJ}^\alpha &=\ii\tau^a_{ij} \int \dd y \, \zeta^{+}_{I}(y)\zeta^{-}_{J}(y)\partial_y f^\alpha(y)\\
&=-\ii( m_I\mathcal{T}_{LIJ}^\alpha -m_J\mathcal{T}_{RIJ}^\alpha)+\ii\tau^a_{ij}\lbrack
 \zeta^{+}_{I}\zeta^{-}_{J}f^\alpha\rbrack_{y_f}\nonumber
\end{align}
This is in agreement with the result \eqref{eq:f-phi-coupling} from the WI
provided the boundary term vanishes.
For the coupling to  broken gauge bosons this follows since
the gauge boson wavefunction $f^a$ at the fixed point vanishes 
for orbifold or Dirichlet BCs. Then 
\eqref{eq:f-phi-coupling} is satisfied independent of the BCs of the fermions.

For the coupling to  unbroken gauge bosons  we have to use the 
consistency condition \eqref{eq:fermion-cons}. 
In the basis where the orbifold transformation of the gauge bosons is
diagonal it implies
\begin{equation}\label{eq:parity-relation}
\tau^{a}_{ij}
(\lambda_i-\lambda_j)=0
\end{equation}
and therefore $\lambda_i=\lambda_j$ if the generators are nonvanishing.
Thus either $\zeta^{+}$ or 
$\zeta^{-}$ vanishes since the left and right handed modes have different
orbifold parity. Therefore \eqref{eq:f-phi-coupling} is satisfied for
orbifold gauge symmetry breaking.
The consistency condition for a more general symmetry breaking by BCs
is that the couplings of the unbroken gauge bosons must connect
fermions with the same BCs.

The relation~\eqref{eq:2phi-w-coupling} can be verified analogously
using the relation among the 
wavefunctions~\eqref{eq:wf-relation}, integrating by 
parts two times and using the equation of motion for the 
KK-wavefunctions~\eqref{eq:kk-dgl}:
\begin{equation}
  \begin{aligned}
    m_{\alpha_y}m_{\beta_y}T^\alpha_{\beta\gamma}&=f^{abc}\int \dd ^N y \, 
    f^\alpha(y)\partial_y f^\beta(y)\partial_y f^\gamma(y)\\
    &=\frac{1}{2}f^{abc}\int d ^N y \,\Bigl[\partial_y^2 f^\alpha(y)f^\beta(y) 
    f^\gamma_{ l}(y)-f^\alpha(y) \partial_y^2f^\beta(y) f^\gamma(y)
    - f^\alpha(y) f^\beta(y)\partial_y^2 f^\gamma(y)\Bigr]\\
    &=\frac{1}{2}(m_{\alpha}^2-m_{\beta}^2-m_{\gamma}^2)g^{\alpha\beta\gamma}
  \end{aligned}
 \end{equation}
The boundary terms occurring in the integration by parts 
are of the form $[\partial_y f^a f^bf^c]$.
As has been shown in~\cite{OS:SR} these terms
vanish 
as long as the wavefunctions of the broken gauge bosons are zero
 on the boundary so both orbifold and general
 Dirichlet BCs \eqref{eq:dirichlet} are 
consistent with unitarity and WIs.

Conditions similar to \eqref{eq:fermi-srs} 
have been discussed in detail for  the gauge boson
SRs in~\cite{Csaki:2003dt,OS:SR} so here we will be brief. 
Performing the sum over the KK-modes using the completeness 
relations for the fermion and gauge boson wavefunctions, the same 
integral over the KK wavefunctions appears in every term and
both equations of~\eqref{eq:fermi-srs} reduce to the Lie algebra
of the generators of the gauge group. For instance, the 
condition~\eqref{eq:fermi-gold-KK} turns into
\begin{equation}\label{eq:complete-simplify}
0=\left([\tau^a,\tau^b]-\ii f^{abc}\tau^c\right)
 \int dy\
 f^\alpha(y)g^\beta(y)\zeta^{+}_{I}(y)\zeta^{-}_{J}(y)
\end{equation}
Note that the unitarity cancellations require to sum over
the KK-towers of both 
fermions and gauge bosons so it is essential that the fermions 
propagate in the bulk.
\section{Consistency of brane localized terms}
\label{sec:brane-masses}
We now turn to the verification of the SRs for theories including 
brane localized terms like~\eqref{eq:dirac-brane}
and~\eqref{eq:brane-mix}.
In section~\ref{sec:gen-bcs}
we review the consistency of BCs with the equations of motion.
In section~\ref{eq:bmt} we discuss the consistency of the BCs corresponding
to brane localized masses, extending the analysis of
section~\ref{sec:orbi-sr}. Brane localized kinetic terms and mixing with
brane fermions are discussed in section~\ref{sec:bkt}.
\subsection{Generalized boundary conditions}\label{sec:gen-bcs} 
There are two approaches to theories on an orbifold: 
in the interval approach  (sometimes called `downstairs' approach)  
one uses continuous fields on the physical
 interval $[0,\pi R]$  while in
 the orbifold (or `upstairs') approach
fields are defined on the circle  $[0,2\pi R[$ and discontinuities
are allowed at the orbifold fixed points $0,\pi R$.
The treatment of brane localized terms like~\eqref{eq:dirac-brane}
and~\eqref{eq:brane-mix} differs in the two approaches.
In the interval approach,
one imposes appropriate BCs at the boundaries 
instead of including the singular terms involving delta functions
in the equations of motion~\cite{Csaki:2003sh}. 
In the orbifold approach, singular terms
are included in the equations
of motion and
ordinary orbifold BCs are imposed at the fixed points.
Nontrivial BCs that determine the discontinuities in the wave
functions and the mass spectrum are derived by
integrating the equations of motion in an infinitesimal interval 
around the fixed points~\cite{Meissner:2002dg,Delgado:2002xf}.

In this section we take the interval point of view and review possible 
BCs consistent with the equations of motion, 
following~\cite{Csaki:2003sh} but focusing on  
models with a left-right 
symmetry in the bulk and BCs corresponding to Dirac masses.
(For a discussion of consistent BCs for fermions in six dimensions
 see~\cite{Dobrescu:2004zi}).

In order to obtain equations of motion without boundary terms, 
the BCs have to be chosen so that the boundary
terms in the variational derivation of the equations of motion vanish. 
To obtain these conditions, we write the kinetic term  
in the symmetric form
\begin{equation}
\mathscr{L}_{\text{kin}}=\frac{1}{2}\ii (\bar\Psi\partial_M \Gamma^M\Psi-
\partial_M \bar\Psi \Gamma^M\Psi)
\end{equation}
where again $\Psi=(\Psi_L,\Psi_R)$.
The boundary terms appearing in the variation of the action are given by
\begin{equation}\label{eq:eom-bc}
  \frac{1}{2}\int \dd^4 x\left[(\delta \bar\Psi)\gamma^5\Psi- 
\bar\Psi\gamma^5\delta\Psi\right]_0^\ell
= \frac{1}{2}\int \dd^4 x
 \left[\delta \chi^\dagger \eta- \delta \eta^\dagger \chi- 
    \chi^\dagger  \delta \eta+ \eta^\dagger \delta \chi
  \right]_{0}^\ell
\end{equation}
We will now impose the BC that the term in brackets vanishes
at each boundary $y_f=0,\ell$.

Of course the simplest solution is to demand that every term 
in~\eqref{eq:eom-bc} vanishes by itself, e.g. by demanding
\begin{equation}
  \eta_L(y_f)=\chi_R(y_f)=0
\end{equation}
Here also the corresponding variations are demanded to vanish.
This corresponds just to the orbifold BCs discussed in the previous section.
A less trivial  solution corresponds to the introduction of a brane Dirac 
mass term~(c.f. appendix~\ref{app:brane-kk}). 
In contrast to the case of Majorana fermions discussed in detail 
in~\cite{Csaki:2003sh}, 
we cannot demand BCs of the form
$\eta_{L/R}\propto \chi_{L/R}$. 
We can, however, choose a BC that mixes $\Psi_L$ and $\Psi_R$:
\begin{equation}\label{eq:mass-bc}
  \begin{aligned}
  \chi_L(y_f)&=-\tan\alpha_{y_f}\chi_R(y_f)\\
  \eta_L(y_f)&=\cot\alpha_{y_f}\eta_R(y_f)  
  \end{aligned}
\end{equation}
so that
\begin{equation}
 \left[ \delta \chi_L^\dagger \eta_L
   +\delta \chi_R^\dagger \eta_R \right]_{y_f}=0
\end{equation}
and so on. 
This generalizes the orbifold BC~\eqref{eq:iso-orbi} that corresponds to 
the special case $\alpha=\frac{\pi}{4}$.
Instead of~\eqref{eq:orbi-masses}
we obtain the mass quantization condition
\begin{equation}\label{eq:KK-bc}
  \frac{\zeta_{L,n}^-(y_f)}{\zeta_{R,n}^-(y_f)}=-\tan \alpha_{y_f}
\end{equation}
Another choice of BCs consistent with the variational principle is
~\cite{Csaki:2003sh}
\begin{equation}\label{eq:bkt-bc}
  \chi(y_f)=\ii\kappa \sigma^\mu \partial_\mu\eta(y_f)
\end{equation}
This BC corresponds to a brane kinetic term (c.f. appendix~\ref{app:bkt}).
Here  the boundary terms~\eqref{eq:eom-bc} vanish since
the operator $\ii\sigma^\mu \partial_\mu$ is hermitian:
\begin{equation}
\int \dd^4 x\,
 \left[ \delta \chi_R^\dagger \eta_R- \delta \eta_R^\dagger \chi_R
 \right]_{y_f}\\
=-\ii\kappa\int \dd^4 x\,
\left[(\partial_\mu \delta \eta_R^\dagger \sigma^\mu ) \eta_R
+\delta \eta_R^\dagger(\sigma^\mu \partial_\mu\eta_R)\right]_{y_f}=0
\end{equation}
A generalization of~\eqref{eq:bkt-bc} appears for mixing with brane localized
fermions (c.f. section~\ref{sec:bkt}).

The advantage of the interval approach adopted in this section
is that discontinuous wavefunctions and the associated
ambiguities are avoided.
In gauge theories, however, consistency with the variational principle 
is not the only consistency requirement.
Assigning BCs at will can violate unitarity or WIs, even if the action 
is gauge invariant and the BCs are consistent with the equations of motion.
A drawback of the interval approach
is that the compatibility of the BCs with the gauge symmetry is
not apparent while
these issues are much more transparent in the equivalent
description in terms of brane localized terms on orbifolds. 
Therefore both approaches will be taken 
into account in the subsequent discussion of the consistency of BCs,
 providing useful cross checks of the results.

As a further alternative to brane located mass terms,
an equivalent description in terms of
Scherk-Schwarz breaking on orbifolds 
has been found for suitable Majorana brane mass 
terms~\cite{Bagger:2001qi}. 
It would be interesting to extend this analysis to
 brane induced  Dirac masses in the context of gauge symmetry
breaking, but this is beyond the scope of this work.
\subsection{Brane localized Dirac masses}
\label{eq:bmt}
We now extend the discussion of section~\ref{sec:orbi-sr} to the more general
BCs~\eqref{eq:mass-bc}. As discussed in appendix~\ref{app:brane-kk} such BCs
arise also from a brane localized mass term. 
We consider the same gauge symmetry  breaking pattern
$SU(2)_L\times SU(2)_R\to
SU(2)_{L+R}$ like in section~\ref{sec:orbifolds} but rather than mixing 
$\Psi_L$ and $\Psi_R$ by an orbifold transformation as in~\eqref{eq:iso-orbi}, 
this is achieved by the mass term~\eqref{eq:dirac-brane}.
While in a realistic model this must be combined with isospin breaking
brane terms at $y=0$, only the BCs at $y=\ell$  are important 
in the subsequent discussion and the consistency of the 
brane localized terms at the other boundary can be discussed separately~(see
section~\ref{sec:bkt}).

For simplicity, we set the bulk masses to zero so the
KK wavefunctions of the left-and right handed wavefunctions are related as
\begin{equation}\label{eq:kk-wfs}
  \begin{aligned}
    \zeta^-_{L,I}(y)&=\zeta^+_{R,I}(y)\equiv\zeta^-_{I}(y)\\
    \zeta^+_{L,I}(y)&=-\zeta^-_{R,I}(y)\equiv\zeta^+_{I}(y)
  \end{aligned}
\end{equation}
(see~\eqref{eq:kk-wfs-twist} for $m=0$). 
The equations of motion~\eqref{eq:bmt-eom-app} imply the equations 
for the KK-wavefunctions:
\begin{equation}\label{eq:bmt-eom}
\partial_5\zeta^\pm_{I}\mp m_I\zeta^\mp_{I}
+\delta(y-\ell)M_D \ell \,\zeta^\mp_{I}=0 
\end{equation}
As derived in appendix~\ref{app:brane-kk}, the KK wavefunctions satisfy the
 BCs~\eqref{eq:KK-bc} with $\alpha=\mathrm{artanh} M_D\ell $ 
(see~\eqref{eq:bmt-bc}).
This agrees with the results of~\cite{Delgado:2002xf} for Majorana mass terms.

Performing the KK-decomposition of the Lagrangian~\eqref{eq:lr-lag}, 
we obtain the interaction terms
\begin{equation}\label{eq:kk-ferm}
\mathscr{L}_{f,KK}=
\frac{1}{\sqrt 2}\bar\psi_I(\fmslash
A^+_\alpha\mathcal{T}^{+\alpha}_{IJ}
+\fmslash A^-_\alpha\mathcal{T}_{IJ}^{-\alpha}\gamma^5)\psi_J
+\bar\psi_I 
(\phi_\alpha^+g_{IJ}^{+\alpha}+\phi_\alpha^-g_{IJ}^{-\alpha}\gamma^5)\psi_J
\end{equation}
with the coupling constants given by
\begin{subequations}
\begin{align}
\mathcal{T}^{\pm\alpha}_{IJ}&=\tau^a_{ij}\int \dd y\,
\Bigl[\zeta^{+}_{I}(y)\zeta^{+}_{J}(y)\pm \zeta^{-}_{I}(y)
\zeta^{-}_{J}(y)\Bigr]
f^{\alpha,\pm}(y)\\
g_{IJ}^{\pm\alpha}&=\mp\ii \tau^a_{ij}\int
\dd y\,\Bigl[\zeta^{+}_{I}(y)\zeta^{-}_{J}(y)\mp
\zeta^{-}_{I}(y)\zeta^{+}_{J}(y)\Bigr]  g^{\alpha,\pm}(y)
\end{align}
\end{subequations}
Following the discussion of the pure orbifold symmetry breaking 
in section~\ref{sec:orbi-sr}, we now verify the unitarity SRs.
The SRs~\eqref{eq:fermi-srs} are satisfied like in the pure orbifold case
by the completeness relation of the KK-wavefunctions.

To check the  
condition~\eqref{eq:f-phi-coupling}, 
let us first employ the interval approach 
where we use the equations of motion without delta-singularities 
and impose the nontrivial BCs~\eqref{eq:KK-bc} instead.
Similarly to~\eqref{eq:f-phi-orbi} we obtain
\begin{multline}\label{eq:f-phi-brane}
m_\alpha g_{IJ}^{\pm \alpha} 
=\pm\ii \tau^a_{ij}\int
\dd y\,\Bigl[\zeta^{+}_{I}(y)\zeta^{-}_{J}(y)\mp
\zeta^{-}_{I}(y)\zeta^{+}_{J}\Bigr]  \partial_y f^{\pm \alpha}(y)\\
=-\ii( m_I\mp m_J)\mathcal{T}_{IJ}^{\pm\alpha}
+\ii \tau^a_{ij}
\left[\left[\zeta^{+}_{I}\zeta^{-}_{J}\mp
\zeta^{-}_{I}\zeta^{+}_{J}\right]f^{\pm \alpha}\right]_{\ell}
\end{multline}
Provided the boundary term vanishes, this is indeed the  
condition~\eqref{eq:f-phi-coupling} 
translated to the vector and axial vector couplings
$\mathcal{T}^\pm=\frac{1}{2}(\mathcal{T}_R\pm\mathcal{T}_L)$ used
in this section.

For the broken gauge bosons $A^-$ the boundary term vanishes since
the wavefunctions of the gauge bosons are zero on the 
boundary. 
For the unbroken gauge bosons $A^+$ the vanishing of the 
boundary term is ensured by the BCs~\eqref{eq:KK-bc}:
  \begin{equation} 
   \left[\zeta^{+}_{I}\zeta^{-}_{J}-
\zeta^{-}_{I}\zeta^{+}_{J}\right]_{y=\ell}=0
  \end{equation}
The same conclusion is reached in the orbifold approach, 
where we impose ordinary orbifold BCs but have
to use the singular equations of 
motion~\eqref{eq:bmt-eom}.
In the coupling to $A^+$~\eqref{eq:f-phi-brane} 
the resulting additional terms at the boundary 
cancel, because they are the same for both $\zeta^+$ and $\zeta^-$.

Evidently, the vanishing of the boundary terms requires that
 the BCs for fermions are consistent with
the breaking of the gauge symmetry. 
For an unbroken left-right gauge symmetry on the brane, the wavefunctions 
of the $A^-$ are nonvanishing. Thus, in spite of consistency with the 
variational principle, the BCs~\eqref{eq:KK-bc} violate the 
unitarity SRs in this case.
This is of course expected since the BCs correspond to brane mass terms
inconsistent with an unbroken left-right symmetry, but this incompatibility
is not apparent in the BC description. 
Similarly, the cancellation of the boundary terms in~\eqref{eq:f-phi-brane}
demands that all components of the isospin doublets
$\Psi_L$ and $\Psi_R$ must satisfy the same BCs~\eqref{eq:KK-bc} so only 
a mass term consistent with the unbroken $SU(2)_{L+R}$ is allowed.
\subsection{Brane localized kinetic terms and mixing with 
brane fermions}\label{sec:bkt}
As discussed in section~\ref{sec:higgsless},
mixing with brane fermions or brane kinetic terms can be used to
obtain a mass splitting among the components of the
 fermion isospin multiplets~\cite{Csaki:2003sh}. 
We now discuss the consistency of this setup.
Brane kinetic terms will be discussed
first, the similar case of mixing with brane fermions is discussed below.
In the example of Higgsless EWSB, the BKTs are added
for the $\Psi_R$ fermions on the brane where a symmetry $SU(2)_L\times U(1)$
is unbroken.
To be consistent with the reduced gauge symmetry on the brane, we
have to add a BKT with a covariant derivative
\begin{equation}\label{eq:bkt}
  \mathscr {L}_{BKT}=\kappa\ii\bar\eta_{R,i}
 (\fmslash\partial\delta_{i,j} +\tau^{a}_{ij}{\fmslash A}^a)
 \eta_{R,j}\,
 \delta(y)
\end{equation}
 where the $A^a$ are 
the unbroken gauge bosons only. In the example of Higgsless EWSB, 
these are the $U(1)$ gauge bosons $B$.
In a more general situation, we can consider a general bulk symmetry group
 $G$ that is broken
to a subgroup $H$ at the boundary by orbifolding or by Dirichlet BCs.
The representation of the fermions
under $G$ can then be decomposed into representations of $H$ and 
we can allow different BKTs on the brane for fermions in
the different representations of $H$. 

Taking the BKT~\eqref{eq:bkt} into account, the equations of 
motion for the KK modes  become
\begin{equation}\label{eq:bkt-wfs}
  \begin{aligned}
    \partial_5\zeta^-_{I}+m_I\left(1+\kappa\delta(y)\right)\zeta^+_{I}&=0\\
    -\partial_5\zeta^+_{I}+m_I\zeta^-_{I}&=0 
  \end{aligned}
\end{equation}
The determination of the BCs and the mass spectrum
 is reviewed in appendix~\ref{app:bkt}.
The BC corresponding to the BKT is given by~\eqref{eq:bkt-bc}
and translates to the
BC for the KK wavefunctions
\begin{equation}\label{eq:bkt-bc-kk}
     \zeta_I^-(y)|_{y=0^+}=\kappa m_I \zeta_I^+(y)|_{y=0^+}
\end{equation}
Again we verify the consistency of this BC with the WIs using
the relation~\eqref{eq:f-phi-coupling}.
The presence of the BKT modifies the coupling of the fermions to the
unbroken gauge bosons to
\begin{equation}\label{eq:bkt-couplings}
  \mathcal{T}^{\alpha}_{RIJ}=\tau^a_{ij}\int \dd y\, \zeta^{+}_{I}(y)\zeta^{+}_{J}(y)f^\alpha(y)(1+\kappa\delta(y)) 
\end{equation}
Similarly to~\eqref{eq:f-phi-orbi} we find after integrating by parts:
\begin{equation}\label{eq:f-phi-bkt}
\begin{aligned}
m_\alpha g_{L IJ}^\alpha 
&= -\ii \int \dd y \, \partial_y\bigl(\zeta^{+}_{I}(y)\zeta^{-}_{J}(y)\bigr)
f^\alpha(y)
+\ii\tau^a_{ij}\lbrack
 \zeta^{+}_{I}\zeta^{-}_{J}f^\alpha\rbrack_{y_f}\\
&=-\ii( m_I\mathcal{T}_{LIJ}^\alpha -m_J\mathcal{T}_{RIJ}^\alpha)
\end{aligned}
\end{equation}
Here the modified coupling constants~\eqref{eq:bkt-couplings} appear 
for the KK-modes of the unbroken gauge bosons in the last expression.
This follows in the interval approach using the 
BC~\eqref{eq:bkt-bc-kk} and the continuous equations of motion. 
In the orbifold-approach it results from the discontinuous equation of 
motion~\eqref{eq:bkt-wfs} and trivial BCs.
For the coupling to the broken gauge bosons, the boundary terms vanish 
since the gauge boson wavefunctions vanish on the boundary.

We therefore have shown that in the presence of unbroken gauge symmetries on
a brane, the modification of the BCs~\eqref{eq:bkt-bc-kk} necessitates the
modification of the couplings of the fermions to the unbroken gauge bosons
according to~\eqref{eq:bkt-couplings}. 
In the orbifold approach, this modification appears naturally by
using covariant derivatives in the BKTs. 

Finally, we turn to mixing with brane fermions that is very
similar to the case of BKTs~\cite{Csaki:2003sh}.
We consider the mixing of the right handed component of the 
five dimensional fermion $\Psi_R=(\chi_R,\eta_R)$ with
a brane localized fermion $\psi=(\chi,\eta)$ at $y=0$ 
via a mass term:
\begin{equation}\label{eq:mix-lag}
  \mathscr{L}_{Mix}=\delta(y)\left[\ii\bar\psi_i (\fmslash \partial\delta_{ij}
    +\tau^a_{ij}\fmslash A^a)\psi_j-\mu\bar\psi_i\psi_i
    -M\ell^{\frac{1}{2}}\left(\eta_{i,R}^\dagger \chi_i
      +\chi_i^\dagger \eta_{R,i}\right)\right]
\end{equation}
Similarly to the BKTs, only a coupling to the unbroken gauge bosons is present.
As discussed in appendix~\ref{app:mix}, the solution of the equation of 
motion of the brane fermions takes the form
\begin{equation}
\psi(x)=\sum_{n} \beta_n\begin{pmatrix}\frac{m_n}{\mu}\chi_{n}(x)\\
                                \eta_{n}(x)\end{pmatrix}    
\end{equation}
with
\begin{equation}\label{eq:brane-sol}
   \beta_n=\ell^{\frac{1}{2}}\frac{M\mu}{(m_n^2-\mu^2)}\zeta^+_{R,n}|_{y=0}
\end{equation}
Here the same spinors $\chi$ and $\eta$ appear as in the KK decomposition
of the bulk fer\-mi\-ons~\eqref{eqref:fermi-kk}. The equations of motion 
and  BCs for the bulk fermions
are the same as in the case of BKTs~\eqref{eq:bkt-bc-kk} with the replacement
\begin{equation}
   \kappa\to \tilde\kappa_I= \frac{\ell M^2}{\mu^2-m_I^2}
\end{equation}
Because the decomposition of the brane fermions~\eqref{eq:brane-sol} 
involves the KK-wavefunctions at the location of the brane, 
the couplings of the fermion KK-modes to the unbroken
gauge bosons gets modified:
\begin{equation}\label{eq:mix-gb-couplings}
  \begin{aligned}
   \mathcal{T}^{\alpha}_{RIJ}&=\tau^a_{ij}\int \dd y\,
   f^\alpha(y)\zeta^{+}_{I}(y)\zeta^{+}_{J}(y)
   \left(1+\delta(y)\tilde\kappa_I\tilde\kappa_J\frac{\mu^2}{\ell M^2}\right)\\
   \mathcal{T}^{\alpha}_{LIJ}&=\tau^a_{ij}\int \dd y\,f^\alpha(y)
   \left(\zeta^{-}_{I}(y)
   \zeta^{-}_{J}(y)+\delta(y) \zeta^{+}_{I}(y)\zeta^{+}_{J}(y)
     \tilde\kappa_I\tilde\kappa_J\frac{m_Im_J}{\ell M^2}\right) 
  \end{aligned}
\end{equation}
Performing the by now usual manipulations to verify~\eqref{eq:f-phi-coupling}, 
we find again
that the singular terms in the equations of motion
(or the nontrivial BCs in the interval approach)
 contribute just the
terms needed to compensate for the changed 
gauge boson couplings:
\begin{multline}\label{eq:f-phi-mix}
-\ii( m_I\mathcal{T}_{LIJ}^\alpha -m_J\mathcal{T}_{RIJ}^\alpha)
=-\ii \int\dd y \, f^\alpha\Bigl[m_I \zeta^{-}_{I}\zeta^{-}_{J}
                  - m_J\zeta^{+}_{I}\zeta^{+}_{J}\left( 1
         + \delta(y)
         \frac{\tilde\kappa_I\tilde\kappa_J}{\ell M^2}(\mu^2-m_I^2)\right)
         \Bigr]\\
=  -\ii \int\dd y \, f^\alpha\Bigl[m_I \zeta^{-}_{I}\zeta^{-}_{J}
                  -m_J \zeta_{R,I}^+ \zeta^{+}_{J}
                  \left(1+\tilde\kappa_J\delta(y)\right)
                 \Bigr]
=m_\alpha g_{L IJ}^\alpha 
\end{multline}
Considering the limit of sending $\mu$ and $M$ to infinity while keeping
$\frac{M}{\mu}\equiv \sqrt{\frac{\kappa}{\ell}}$ fixed~\cite{Csaki:2003sh} we have
$\tilde \kappa_n\to\kappa$ and recover the same BCs and equations of motion as
for BKTs. As required by this analogy, the modification in $\mathcal{T}_L$ in 
\eqref{eq:mix-gb-couplings} vanishes in this limit.
Again we have found that modified BCs are consistent with the gauge symmetry only
if they correspond to a boundary term invariant under the reduced gauge
symmetry and the peculiar form of the 
additional terms in~\eqref{eq:mix-gb-couplings} enforced by gauge symmetry
has played an essential role in verifying the consistency.

\section{Summary and conclusions}
We have applied the sum rules obtained from tree 
unitarity~\cite{LlewellynSmith:1973} 
and  Ward identities~\cite{OS:SR} to discuss the consistency
of higher dimensional mechanisms for fermion masses without
Higgs bosons.
In section~\ref{sec:wi-results}
we have introduced an orbifold mechanism to obtain Dirac masses for 
five dimensional fermions in a chiral theory and have checked the
SRs for general orbifold gauge theories involving bulk fermions.
Similar to a pure KK gauge theory, the unitarity sum rules
are satisfied by interlacing cancellations among the KK-states 
of both bulk fermions and gauge bosons.

To obtain a mass splitting 
among the components of the isospin doublets in the SM, the orbifold
mechanism is not sufficient and  generalized boundary conditions
corresponding to brane localized mass and kinetic terms 
and mixing with brane fermions are employed in models of Higgsless EWSB.
In section~\ref{sec:brane-masses}
the consistency of these  boundary conditions
has been discussed.
While the approach of~\cite{Csaki:2003sh} to impose BCs consistent
with the variational principle at the boundaries of an interval
avoids ambiguities from discontinuous wavefunctions, the consistency
of the BCs with gauge symmetries is more transparent in the 
equivalent description in terms of brane localized terms. 
We have found that indeed  only boundary conditions corresponding to brane 
localized mass and kinetic terms respecting the reduced
gauge symmetry on the brane are consistent with unitarity
and WIs so they can be considered as soft symmetry breaking. 

Apart from the models of Higgsless EWSB that have served
 as example in this work, 
the consistency of the picture of explicit but soft symmetry breaking by
brane localized terms invariant under a reduced gauge symmetry is also  
expected to be important for Higgsless
models of gauge unification in higher dimensions.

Another interesting question that is left for future work is the
possible description of brane localized mass terms in gauge theories 
in terms of Scherk Schwarz breaking~\cite{Bagger:2001qi}.
\subsection*{Acknowledgments}
I thank Thorsten Ohl for  reading the manuscript.
This work has been
supported by the Deutsche
Forschungsgemeinschaft through the Gra\-du\-ier\-ten\-kolleg
`Eichtheorien' at  Johannes Gutenberg University, Mainz.
\appendix
\numberwithin{equation}{section}
\section{Effective lagrangians for Kaluza-Klein modes}\label{app:kk}
\subsection{Kaluza-Klein decomposition of gauge bosons}\label{app:kk-gauge}
In this appendix we set up our notation for the KK lagrangian and 
decomposition of the gauge bosons, following the general 
higher dimensional case discussed in~\cite{OS:SR}.

 The 5 dimensional Yang Mills lagrangian is given by
\begin{equation}
   \mathscr{L}_{5}=-\frac{1}{4} F^a_{AB}(x,y)F^{a,AB}(x,y)
\end{equation}
with the field strength
\begin{equation}
  F^a_{AB}(x,y)=\partial_A A^a_B(x,y)
     - \partial_B A^a_A(x,y) + f^{abc}A^b_A(x,y)A^c_B(x,y)
\end{equation}
Here we include the higher dimensional gauge coupling $g_5$ in the
structure constants.
We use a `mostly minus' metric $g_{AB}=\text{diag}(\eta_{\mu\nu},-1)$.
The KK decomposition of the gauge fields is introduced as 
\begin{equation}\label{eq:kk-decomp}
A^a_A(x,y)=
\begin{pmatrix}
  A^a_\mu(x,y)\\ -\phi^a(x,y)
\end{pmatrix}
=\sum_{ n}\begin{pmatrix}
 f^a_{ n}(y) A^a_{ n,\mu}(x)\\ -g^a_{ n}(y) \phi^a_{ n}(x)
\end{pmatrix}
\end{equation}
The sign of the scalar component 
is chosen because of compatibility with our
conventions for the WIs used in~\cite{OS:SR}.
The wavefunctions $\rho=f,g$ satisfy the differential equation
\begin{equation}\label{eq:kk-dgl}
\partial_y^2 \rho_{ n}^a(y)=-{m^a_{ n}}^2\rho^a_{ n}
\end{equation}
They are taken as orthonormal and satisfying a completeness relation: 
\begin{align}
\label{eq:complete}
  \int\dd ^N y\, \rho^a_{ n}(y) \rho^a_{ m}(y)&=\delta_{ n, m}\\
\sum_{ n} \rho^a_{ n}(x)\rho^a_{ n}(y)&=\delta(y-x)
\end{align}
In these equations the  group indices are not summed over.
To diagonalize the interaction of the vector and scalar components
we choose~\cite{OS:SR}
\begin{equation}\label{eq:wf-relation}
\partial _y f_{ n} = m_{n} g_{ n}\quad,\quad 
\partial _y g_{ n}= -m_{n} f_{ n}
\end{equation}
This relations can be used to obtain the effective four dimensional
Lagrangian of the KK modes. To simplify our notation, we introduce 
a multi-index notation with $(a,i)\equiv \alpha$.
The cubic interaction terms relevant for the SRs discussed in section
~\ref{sec:wi-results} are found as
\begin{equation}
\label{eq:kk-int}
  \mathscr{L}^{KK}_{\text{cubic}} =
    - g^{\alpha\beta\gamma}\partial_\mu A_\nu^\alpha A^{\beta,\mu}A^{\gamma,\nu} 
    - \frac{1}{2}T^{\alpha}_{\beta\gamma}\, A^{\alpha,\mu}
            \phi^{\beta}\overleftrightarrow{\partial_\mu} \phi^\gamma
            +\dots
\end{equation}
where the coupling constants are given by
\begin{subequations}
\label{eq:kk-couplings}
\begin{align}
\label{eq:g_abc}
  g^{\alpha \beta\gamma}
    &= f^{abc}\int\!\dd^N y\, f^\alpha (y)f^\beta(y) f^\gamma (y)\\
  T^{\alpha}_{\beta\gamma}
    &= f^{abc}\int\!\dd^N y \, f^\alpha (y)g^\beta (y)g^\gamma (y)
\label{eq:t_abc}
\end{align}
\end{subequations}
The complete lagrangian and the coupling constants
in an arbitrary number of dimensions have been given in~\cite{OS:SR}.
\subsection{KK-decomposition for Fermions}\label{app:kk-ferm}
We now introduce our notation for fermions  on a 5-dimensional 
orbifold~\cite{Georgi:2000,Papavassiliou:2001be}.
The Lagrangian is taken 
as\footnote{We use a 4-component notation for the spinors with 
the notation of \cite{Peskin:1995} i.e. 
$\gamma^\mu=\begin{pmatrix}0&\sigma^\mu\\\bar\sigma^\mu&0\end{pmatrix}$
 and $\gamma^5=\begin{pmatrix}-1&0\\0&1\end{pmatrix}$}:
\begin{equation}\label{eq:fermi-lag}
\mathscr{L}_f=\bar \Psi_i(x,y)(\ii \partial_M \Gamma^M-m_i(y))\Psi_i(x,y) 
+ \bar\Psi_i(x,y) \tau^a_{ij}\Gamma^M  A^a_M(x,y)\Psi_j(x,y) 
\end{equation}
with the 5-dimensional gamma-matrices
\begin{equation}
\Gamma^\mu=\gamma^\mu \,,\, \Gamma^5=\ii \gamma^5
\end{equation}
and where the $\tau^a$ are the generators of the gauge group in the representation 
of the fermions. The mass function must be odd under the orbifold transformations~\eqref{eq:orbi-trans}. 

The resulting equation of motion for the free fermion fields is 
\begin{equation}
(  \ii\fmslash\partial-\gamma^5\partial_5-m_i(y))\Psi_i(x,y)=0
\end{equation}
We introduce the KK decomposition for the left-and righthanded components:
\begin{equation}
\Psi(x,y)=
\begin{pmatrix}
  \chi(x,y)\\\eta(x,y)
\end{pmatrix}
=
\sum_{n}\begin{pmatrix}\chi_{n}(x)\zeta^-_{n}(y)\\
                                   \eta_{n}(x)\zeta^+_{n}(y)\end{pmatrix}
\end{equation}
where  $\eta_n,\chi_n$ are 4-dimensional right- and  lefthanded spinors
that satisfy the appropriate Dirac equations:
\begin{equation}\label{eq:4d-dirac}
\ii\sigma^\mu \dmu\eta_{n}-m_{n}\chi_{n}=0\quad,\quad
\ii\bar\sigma^\mu \dmu\chi_{n} -m_n\eta_{n}=0
\end{equation}
The KK wavefunctions satisfy the equations of motion
\begin{equation}\label{eq:ferm-kk-eom}
(\mp\partial_y -m(y))\zeta^\pm_n=-m_n\zeta_n^{\mp}
\end{equation}
and completeness and
orthogonality relations similar to~\eqref{eq:complete} hold.

Inserting the KK decompositions into the Lagrangian and integrating over the 
fifth dimension  
results in the interaction lagrangian:
\begin{equation}
\mathscr{L}_{fKK}=
\bar\psi_I\fmslash
A_\alpha(\mathcal{T}^{\alpha}_{LIJ}(\tfrac{1-\gamma^5}{2})
+\mathcal{T}_{RIJ}^\alpha(\tfrac{1+\gamma^5}{2})\psi_J
+\bar\psi_I \phi_\alpha
(g_{L IJ}^\alpha(\tfrac{1-\gamma^5}{2})+g_{R IJ}^{\alpha}(\tfrac{1+\gamma^5}{2}))\psi_J
\end{equation}
where we have used a multi-index notation with
 $(n,i)\equiv I$ and defined the 4 dimensional
Dirac spinors $\psi_I=(\chi_I,\eta_I)$. Since only the left or righthanded
component possesses a zero mode, either $\chi_{0,i}$ or $\eta_{0,i}$ vanishes, 
we take this as understood.

The coupling constants are given by 
\begin{subequations}\label{eq:ferm-couplings}
\begin{align}
\mathcal{T}^{\alpha}_{R/LIJ}&=\tau^a_{ij}\int\dd y\, \zeta^{\pm}_{I}(y)\zeta^{\pm}_{J}(y)f^\alpha(y)\\
g_{L/R IJ}^\alpha&=\pm\ii \tau^a_{ij}\int\dd y\,\zeta^{\pm}_{I}(y)\zeta^{\mp}_{J}(y)g^\alpha(y)
\end{align}
\end{subequations}
\section{Boundary conditions for brane localized terms}
\subsection{Brane mass terms}
\label{app:brane-kk}
The equations of motion obtained in the presence of the brane mass 
term~\eqref{eq:dirac-brane} are
given by
\begin{equation}\label{eq:bmt-eom-app}
\begin{aligned}
\ii\sigma^\mu \dmu\eta_L + \partial_5\chi_L
-\delta(y-\ell)l M_D \chi_R&=0\\
\ii\bar\sigma^\mu \dmu\chi_L -\partial_5\eta_L
-\delta(y-\ell)l M_D \eta_R&=0 
\end{aligned}  
\end{equation}
and the same equations with left and right handed fermions exchanged. 
The solutions have been found
in~\cite{Delgado:2002xf} in the context of 
 Majorana brane masses. 
Here we discuss only the BCs and the mass spectrum, 
following~\cite{Meissner:2002dg}.
Using the definition of
the wavefunctions~\eqref{eq:kk-wfs} and the equations 
of motion~\eqref{eq:bmt-eom}, we obtain a decoupled differential equation
for the ratio of the two 
wavefunctions $t_m(y)=\frac{\zeta^+}{\zeta^-}$:
\begin{equation}\label{eq:t-dgl}
  \partial_5 t_n(y)
 =(1+t_n^2)m_n-(1-t_n^2)\delta(y-\ell)M_D \ell
\end{equation}
This equation can used to determine the BCs of the wavefunctions, e.g. by
integrating over a symmetric interval $[\ell -\epsilon,\ell+\epsilon]$ in the
orbifold approach. In the interval approach on may integrate over  
$[\ell -\epsilon,\ell]$ and define 
$\int_0^\ell\delta(y-\ell)=\frac{1}{2}$~\cite{Meissner:2002dg}. 
Here we follow~\cite{Csaki:2003sh} and displace the delta function
to $y=\ell -\frac{\epsilon}{2}$ and impose the same BCs at both boundaries:
\begin{equation}\label{eq:brane-bcs}
  \partial_y\zeta_n^-(y)|_{y=0,\ell}=0\quad, \quad\zeta_n^+(y)|_{y=0,\ell}=0
\end{equation}
Integrating over the interval $[\ell -\epsilon,\ell]$ results in the BCs
\begin{equation}\label{eq:bmt-bc}
 t_{n}(y)|_{y=\ell^-}=
 \mathrm{artanh} M_D\ell 
\end{equation}
i.e. \eqref{eq:KK-bc} with $\alpha=\mathrm{artanh} M_D\ell  $.
Introducing an ansatz compatible with the BCs~\eqref{eq:brane-bcs} at $y=\ell $
\begin{equation}\label{eq:brane-wfs}
  t_n(y)=\tan(m_n(y-\ell)-\varphi_\ell (y))
\end{equation}
the differential equation~\eqref{eq:t-dgl} reduces to:
\begin{equation}\label{eq:phi-dgl}
\partial_y \varphi_\ell(y)
  =\frac{(1-t_n^2)}{(1+t_n^2)}\delta(y-\ell)M_D \ell 
\end{equation}
From this, we obtain in agreement with~\cite{Delgado:2002xf}:
\begin{equation}
  \varphi_\ell (y)=\arctan(\tanh\delta_\ell\epsilon(y-\ell))
\end{equation}
where $\delta_\ell=\frac{M\ell }{2}$ and $\epsilon(y)$ is the sign function
with periodicity $2\pi$
\begin{equation}
  \epsilon(y)=  
 \begin{cases}
  -1,&  -\pi R\leq y < 0\\
  0,& y=0\\
  1,& 0< y\leq \pi R
  \end{cases}
\end{equation}
While the BCs at $y=\ell $ are satisfied by construction, the BCs at $y=0$
yield the mass quantization condition
\begin{equation}\label{eq:brane-masses}
  m_n=\frac{n}{R}-\varphi_\ell(0)=\frac{n}{R}+\arctan(\tanh\delta_\ell))
\end{equation}
\subsection{Brane kinetic terms}\label{app:bkt}
The equations of motion in the presence of boundary kinetic terms are
given by
\begin{equation}
  \begin{aligned}
    \ii\left(1+\kappa\delta(y)\right)\sigma^\mu \dmu\eta + \partial_5\chi
    &=0 \\
    \ii\bar\sigma^\mu \dmu\chi -\partial_5\eta&=0
   \end{aligned}
\end{equation}
To solve these equations, we will impose the BCs
\begin{equation}
  0=\chi(y)|_{y=0,\ell}\qquad 0=\partial_y\eta(y)|_{y=0,\ell}
\end{equation}
and locate the brane an infinitesimal
distance away at $y=\frac{\epsilon}{2}$.
The solution for the KK wavefunctions can be found in~\cite{Dvali:2001gm}.
Here we follow the same approach as in
the case of brane masses to determine the mass spectrum. 
The equation for $t_n$ is given by
\begin{equation}
  \partial_5 t_n=m_n(1+t_n^2)+\kappa m_nt_n^2 \delta(y)
\end{equation}
From this we determine the BC at $y=0^+$ as
\begin{equation}
  t_n(y)|_{y=0^+}=-\frac{1}{\kappa m_n}
\end{equation}
in agreement with~\eqref{eq:bkt-bc}.
In this case, an ansatz compatible with the BCs is given by
\begin{equation}
  t_n(y)=-\cot(m_n y-\varphi_0 (y))
\end{equation}
and the resulting equation for $\varphi_0$ reads
\begin{equation}
\partial_y \varphi_0(y)
  =\frac{t_n^2}{(1+t_n^2)}\delta(y) \kappa m_n 
\end{equation}
We obtain
\begin{equation}
   \varphi_0(y)=\arctan(-\tfrac{1}{2}\kappa m_n\epsilon(y))
\end{equation}
The BCs at $y=\ell$ result in the mass quantization condition
\begin{equation}
  \tan m_n\ell =-\frac{\kappa m_n}{2}
\end{equation}
also found from the explicit solution for the 
wavefunctions~\cite{Dvali:2001gm}. 
\subsection{Mixing with brane fermions}
\label{app:mix}
In the presence of mixing
of brane and bulk  fermions~\cite{Csaki:2003sh}, 
the equations of motion resulting from~\eqref{eq:mix-lag} read
\begin{equation}\label{eq:mix-eoms}
  \begin{aligned}
    \ii\sigma^\mu \dmu\eta_R + \partial_5\chi_R
    -\delta(y)\ell^{\frac{1}{2}} M \chi&=0\\
    \ii\bar\sigma^\mu \dmu\chi_R -\partial_5\eta_R
    &=0 \\
 \ii\sigma^\mu \dmu\eta -\mu\chi &=0\\
\ii\bar\sigma^\mu \dmu\chi-\mu\eta-\ell^{\frac{1}{2}} M\eta_R|_{y=0}&=0
  \end{aligned}
\end{equation}
The first equation implies the BC
\begin{equation}\label{eq:mix-bc1}
   \chi_R(y)|_{y=0^+}
=\ell^{\frac{1}{2}} M \chi
\end{equation}
We decompose the brane fermions as 
\begin{equation}
  \psi(x)=\sum_{n}\begin{pmatrix}\alpha_n\chi_{n}(x)\\
                                 \beta_n\eta_{n}(x)\end{pmatrix}
\end{equation}
 The coefficients $\alpha_n$ and $\beta_n$ are fixed by the last
two equations of~\eqref{eq:mix-eoms}:
\begin{equation}
\alpha_n=\frac{m_n}{\mu}\beta_n\qquad,\qquad
   \beta_n=\ell^{\frac{1}{2}}\frac{M\mu}{(m_n^2-\mu^2)}\zeta^+_{R,n}|_{y=0}
\end{equation}
Using the usual KK-decomposition~\eqref{eqref:fermi-kk}, we then
obtain the equations of motion for the KK-modes
\begin{equation}
  \begin{aligned}
    \partial_5\zeta_{R,n}^-&
    =-m_n\left(1+\tilde\kappa_n\delta(y)\right)\zeta_{R,n}^+\\
    \partial_5\zeta_{R,n}^+&=m_n\zeta_{R,n}^-
  \end{aligned}
\end{equation}
with
\begin{equation}
  \tilde\kappa_n= \frac{\ell M^2}{\mu^2-m_n^2}
\end{equation}
Thus the  wavefunctions satisfy the same equations 
of motion~\eqref{eq:bkt-wfs} and hence also the same  BCs~\eqref{eq:bkt-bc-kk}
as in the case of BKTs with the replacement 
$\kappa\to\tilde\kappa_n$.


\end{document}